# Effect of the disorder in graphene grain boundaries: A wave packet dynamics study

Péter Vancsó[1,4*], Géza I. Márk[1,4], Philippe Lambin[2], Alexandre Mayer[2], Chanyong Hwang[3,4], and László P. Biró[1,4]

[1]Institute of Technical Physics and Materials Science, Centre for Natural Sciences, H-1525 Budapest, P.O. Box 49, Hungary

[2] Department of Physics of Matter and Radiation, University of Namur, 61, Rue de Bruxelles, B-5000 Namur, Belgium

[3]Center for Nano-characterization, Division of Industrial Metrology, Korea Research Institute of Standards and Science, Yuseong, Daejeon 305-340, Republic of Korea

[4]Korean-Hungarian Joint Laboratory for Nanosciences, H-1525 Budapest, P.O. Box 49, Hungary

**Abstract**

Chemical vapor deposition (CVD) on Cu foil is one of the most promising methods to produce graphene samples despite of introducing numerous grain boundaries into the perfect graphene lattice. A rich variety of GB structures can be realized experimentally by controlling the parameters in the CVD method. Grain boundaries contain non-hexagonal carbon rings (4, 5, 7, 8 membered rings) and vacancies in various ratios and arrangements. Using wave packet dynamic (WPD) simulations and tight-binding electronic structure calculations, we have studied the effect of the structure of GBs on the transport properties. Three model GBs with increasing disorder were created in the computer: a periodic 5-7 GB, a "serpentine" GB, and a disordered GB containing 4, 8 membered rings and vacancies. It was found that–for small energies ($E = EF \pm 1\text{eV}$) the transmission decreases with increasing disorder. Four membered rings and vacancies are identified as the principal scattering centres. Revealing the connection between the properties of GBs and the CVD growth method may open new opportunities in the graphene based nanoelectronics.

---

[*] Corresponding author. Email-address: vancso.peter@ttk.mta.hu (P. Vancsó) Tel.:+361-3922222/1316.

1. Introduction

Investigating the different lattice defects and understanding their effects on the charge transport in graphene [1] plays a central role both in the theoretical [2,3] and experimental [4,5] graphene research. Chemical vapor deposition (CVD) on various metal surfaces [6,7] is an effective method to produce graphene samples for practical applications [8,9]. However, recent AFM [10] and STM [11] measurements have shown that especially in the samples grown on Cu foil the graphene lattice breaks into many single-crystal regions with grain boundaries (GBs) between the domains and the presence of GBs may substantially affect the remarkable properties of the perfect graphene lattice [12].

The transport properties of the GBs can be significantly different depending on their detailed geometry. *Ab-initio* transport calculations for periodic pentagon-heptagon dislocation cores [13] predict high transparency or perfect reflection depending on the different misorientation angles of the two grains. Taking into account more complex forms of GBs without periodicity [11,14], Tuan et al. [15] have performed real-space (order-N) quantum transport calculations to determine the Kubo-Greenwood conductivity of polycrystalline graphene samples. They found a very simple scaling law between the semiclassical conductivity and the average grain size. The results also highlighted the strong influence of the different stitching between the grains on the transport properties as it was experimentally also observed [16]. To improve the conductivity of the CVD produced graphene by fine-tuning the growing parameters it is necessary to understand the detailed geometry of the GBs and their effects on the transport.

In our previous calculations the dynamics of the electrons represented by wave packets (WPs) was examined in nanotubes [17] and in graphene [18,19] by the wave packet dynamical method (WPD) [20]. In those papers we presented a detailed analysis of the quantum tunnelling and the charge spreading phenomena in the case when the electron was injected from a simulated STM tip onto the carbon nanostructures. By using a similar geometry, in this paper we present WPD transport calculations on different types of GBs: a periodical pentagon-heptagon GB, a non-periodical "serpentine-like" GB and a non-periodical disordered GB, which contains different defects and non-hexagonal carbon rings. In the first and the second geometry, the C atoms preserve their $sp^2$ symmetry, but the disordered geometry has two-coordinated C atoms as well. By the detailed study of energy dependent wave functions, we were able to identify the main scattering centres in the modelled GBs. Our results give a deep insight into the transport phenomenon of the graphene GBs and can also provide valuable information for practical applications.

## 2. Computational methods

In the CVD growth conditions the Smoluchowski-type ripening may occur [7], where the coalescence of the growing large 2D islands of graphene governs the kinetic pathway. We modelled this process by a Monte-Carlo like method (described in Ref. [21]) to construct the serpentine and disordered GBs: (1) Two graphene cells are generated on the computer with different orientations and one of them is translated so as to overlap the other cell. The C atom sites that belong to the overlapping region (1nm in width) are emptied. (2) Each empty site receives a weight: 0 when its three nearest neighbors are empty, 1 or 2 if respectively one or two nearest neighbor sites are occupied by C atoms. (3) An empty site is selected randomly with a probability proportional to its filling weight. Then it is filled with a C atom which assures that the two grains grow towards the grain boundary. Whenever an empty site lies close to the C atoms of the other grain, its filling weight is reset to 0 to prevent the growing grains to interpenetrate. (4) We continue the filling procedure until all the remaining empty sites have received a zero filling weight. (5) From different randomly chosen positions of the starting grains, the full procedure is repeated several times. The finally selected structure has the smallest mean square deviation of the inter-grain bond lengths with respect to the nominal distance 0.142 nm. (6) At last, the selected structure is relaxed using the Tersoff-Brenner potential while keeping the structure flat, assuming it is bound to a substrate surface.

In order to calculate the transport properties of the GBs we used the WPD method. The dynamics of the electrons wave function $\psi(\vec{r},t)$ is computed from the time dependent 3D Schrödinger equation using the split operator Fourier-transform method [22].

$$\psi(\vec{r}, t + \Delta t) = e^{-i\hat{H}\Delta t} \psi(\vec{r},t) \qquad (1)$$

$$e^{-i(\hat{K}+\hat{V})\Delta t} = e^{-i\hat{K}\Delta t/2} e^{-i\hat{V}\Delta t} e^{-i\hat{K}\Delta t/2} + O(\Delta t^3) \qquad (2)$$

where the Hamiltonian consists of two parts: the kinetic $\hat{K}$ and the potential $\hat{V}$ energy operators. The effect of the kinetic energy propagator $\exp(-i\hat{K}\Delta t/2)$ is given in $k$ space by multiplicating the momentum space wave function $\phi(\vec{k},t)$ by $\exp(-i|\vec{k}|^2 \Delta t/4)$, and the potential energy propagator is a simple multiplication with $\exp(-iV(\vec{r})\Delta t)$ for local potentials. In our potential model the metallic STM tip is represented by a jellium potential (see Ref. [19] for its parameters). For the graphene, we used a local one electron pseudopotential [23] matching the *ab-initio* band structure of graphene sheet π electrons.

$$V_{graphene}(\vec{r}) = \sum_{j=1}^{N}\sum_{i=1}^{3} A_i e^{-a_i|\vec{r}-\vec{r}_j|^2} \qquad (3)$$

where $\vec{r}_j$ denote the atomic positions and N is the number of the atoms. The $A_i$ and $a_i$ coefficients are given in Ref. [23]. Absorbing boundary conditions are realized by a drain potential around the presentation box. With this method the time evolution of the $\psi(\vec{r},t)$ wave function can be determined from any arbitrary initial wave function $\psi(\vec{r},t=0)$. The initial wave function $\psi(\vec{r},t=0)$ was a Gaussian wave packet with $\Delta x, y, z = 0.37$ nm real space width which is considerably larger than the width of the STM tip-graphene tunnelling channel 0.1nm. The Gaussian WP was launched from the tip bulk towards the apex of the tip with momentum equal to the Fermi momentum: $\vec{k} = (0,0,-k_F)$.

From the time-dependent wave function we can obtain all the measurable quantities such as the $\rho(\vec{r},t) = |\psi(\vec{r},t)|^2$ probability density, the $\vec{j}(\vec{r},t) = \text{Im}(\psi^* \cdot \nabla \psi)$ probability current density, etc. Plane integration of $\vec{j}(\vec{r},t)$ along a selected measurement plane gives the probability current $I(t)$. To study the dynamics in the energy domain we performed time-energy Fourier transform [24]. The time dependent wave function can be written as a sum of the eigenstates:

$$\psi(\vec{r},t) = \sum_n A_n \psi(\vec{r},E_n) e^{-iE_n t} \qquad (4)$$

where $A_n$ are the complex superposition coefficients. Applying the t-E Fourier transform to this gives

$$\psi(\vec{r},E) = \int_0^{T_{max}} \psi(\vec{r},t)w(t)e^{iEt}dt = \sum_n A_n \psi(\vec{r},E_n)\delta(E-E_n) \quad (5)$$

where $w(t)$ is a window function which is used to minimize the effect of the finite $T_{max}$ integration time. The energy resolution of the $\psi(\vec{r},E)$ wave function can be increased to any desired accuracy by increasing the $T_{max}$ integration time. A 0.1 eV accuracy was used in our calculation.

### 3. Results and discussion

3.1 Transport results of the GBs

Experimental STM and HRTEM measurements on graphite [25] and graphene [11,26] explored that the perfect hexagonal lattice contains several types of GBs depending on the preparation process. We modelled three different types of GBs along the way described in Sec. 2. Fig. 1a shows a periodical pentagon-heptagon (5-7) GB [25] which has a small periodicity (0.65nm) and very low formation energy. Breaking the periodicity and the rectilinearity of the 5-7 GB the structure became a more realistic, serpentine GB (Fig. 1b) [26]. All but one C atoms remain three-coordinated in this serpentine GB, preserving the sp$^2$ hybridization. Finally, taking a step towards the poorly connected graphene grains [11,14], we generated an aperiodic structure with different non-hexagonal carbon rings from squares to nonagons and two-coordinated C atoms (Fig. 1c). This disordered structure is a large angle GB having a similar misorientation angle as the periodical and the serpentine GBs. These three cases allow us to separate the effect of the different factors such as the periodicity or the C atom coordination number on the transport values.

In our WPD calculations the electron WP is injected from the simulated STM tip situated over a C atom 2.5 nm away from the GB line on the right-hand side grain. After the WP has reached the tip apex from inside tip bulk, it begins to tunnel to the graphene, then spread along the surface. When the tunneled WP hits the GB, part of it is reflected back into the right-hand side grain, but part of it is transmitted through the barrier into the left-hand side grain. Transport calculations are based on the energy dependent probability current $I(E)$ calculated from the energy dependent wave functions $\psi(\vec{r}, E)$ (See Section 2). The transmission coefficient of the barrier is defined as $T(E) = I_{trans}(E)/I_{inc}(E)$, where the measurement plane is perpendicular to the graphene surface and situated 0.5 nm from the GB line on the left-hand side grain. The numerically calculated 3D wave function, however, contains an interference of the incident and reflected wave functions. Therefore, we performed reference calculations without a GB with the same STM tip position in order to separate the incident current from the reflected one. According to the transmission definition $T(E)$ values are between zero and one at all energies. In the case of the pristine graphene $T(E)$ is a constant function takes the value one.

The transmission functions of three modelled GBs at the above-mentioned STM tip geometry are shown in Fig. 2. We observed transmission peaks at similar energy positions for the periodical 5-7 and for the serpentine GB. At these energies (±0.4eV, ±1eV), the periodical 5-7 GB has local density of states (LDOS) maxima originating from the pentagons and heptagons. These higher transmission values can be explained by an increased number of conductance channels according to the Landauer transport theory [21]. The same energy positions in the transport values of the two GBs indicate that the high ratio of the pentagons and heptagons located on a curved line without periodicity has similar electronic structure as the periodical one. Moreover, we can conclude that the high transparency of the periodical 5-7 GB is not affected by breaking the periodicity, as shown by the transmission values for the serpentine GB. In contrast to the previous geometries, the disordered GB has substantially suppressed transmission values between -2eV and +2eV. The presence of the different polygonal rings and two-coordinated atoms may have dramatic effect both on the electronic structure and on the transport values. Surprisingly, at high and low energy positions $E = \pm \gamma_0$, which correspond to Van Hove singularities in the perfect graphene π DOS, all transmission functions have equivalent but small values. At these energy levels the electrons in graphene have a highly anisotropic spreading along the zig-zag directions [19]. This effect, called trigonal warping, is based on the peculiar band structure of graphene, and is strongly connected with the lattice orientation. Hence the transmission between the graphene grains significantly depends on their misorientation angle at the energy region where the trigonal warping is dominant [21]. In our GB structures, the misorientation angles are roughly the same, which explains the similar transport values at $E = \pm \gamma_0$.

The main scattering centers of the disordered GB can be identified by the investigation of the energy dependent wave functions. The calculated $\rho(\vec{r}, E)$ probability density image (Fig. 3) illustrates the spreading pattern of an electron coming from the STM tip with a well defined energy value. Below the STM tip, localized states occur [19] due to the multiple scattering process. In order to display both the strong localized states below the tip apex and the much smaller density maxima in the GB region in the same image, we applied different color scales in the near region (inside a circle of 1.5 nm radius) and in the far region (outside the circle of 1.5 nm radius). Two main scattering centers have been observed on the GB around the Fermi energy: i) the four membered carbon rings (Fig. 3a) and ii) the vacancy type of defects containing 3 two-coordinated C atoms (Fig. 3b). The strong localization and the enhanced reflection of these defects explain the decreased conductivity of the disordered GB.

### 3.2 Electronic structure calculation of the disordered GB

The WPD results for the disordered GB are confirmed with tight-binding (TB) electronic structure calculations. We used a $\pi$ tight-binding Hamiltonian with first-neighbor hopping interactions $V_{ij} = -\gamma_0 (d_{CC}/d_{ij})^2$ where $\gamma_0 = +2.7 eV$ and $d_{ij}$ is the distance between the atoms i and j. The LDOS was computed with the recursion method [27]. Fig. 4 shows the LDOS functions of the corresponding defects in the GB. Finite LDOS values were observed around the Fermi energy for the four membered carbon ring (nr 4.) and the vacancy type of defect (nr 3.). Between the two previous defects a pentagon (nr 5.) and a nonagon (nr 2.) can be seen which also have finite LDOS values around the Fermi energy. However, another nonagon (nr 1.) in a different position does not have peak at the Fermi energy which reveals that the origin of the LDOS peaks at the Fermi energy for defects nr 2. and nr 5. comes from the four membered carbon ring and the vacancy type defect. The similar energy peak positions of nr 5. and nr 2. compared to nr 3. and nr 4. also confirmed this assumption. The effect of the local environment on the LDOS functions can be shown by comparing the two four membered carbon rings (nr 4. and nr 4'.) of the GB structure (Fig. 5b). The peak positions are only slightly different, but the intensity of the LDOS functions differ considerably. The LDOS maximum of the other four membered carbon ring (nr 4'.) is located below the Fermi energy (EF - 0.3 eV) illustrated in Fig. 5a.

In the case of the periodical 5-7 and serpentine GB the LDOS maxima coincide with the transmission maxima. However, in the disordered GB the LDOS peaks around the Fermi energy cause transmission minima. The apparent contradiction can be resolved by taking into account that the vacancy type defects and the four membered carbon rings are sharp lattice defects, therefore they can cause intervalley scattering of the electrons, leading to weak localization [28] and transport reduction. Our observation is in good agreement with other transport calculations [15] where, despite the finite LDOS around the Fermi energy, the transport is remarkably suppressed.

3.3 Effect of the vacancies on the transport

In the previous section we identified two main types of defects that act as scattering centers in the disordered GBs: the four membered carbon rings and the vacancy type defects. The four membered carbon rings appear only in large angle GBs [29], while the vacancy type defects with two-coordinated carbon atoms can better describe poorly connected graphene grains [15]. We carried out WPD calculations on a graphene lattice with increasing number of vacancies along a curved line in the 7.68 nm presentation window. The vacancies (6, 10, 14) are created on the perfect graphene lattice with equal ratio on the two sublattices A and B (Fig. 6). With the help of these geometries we are able to investigate solely the effect of the vacancies on the transport. The integrated values of the transmission functions $\int_{-1eV}^{+1eV} T(E)dE$ for the low-energy carriers (EF ±1 eV) in the cases of the different number of vacancies are shown in Fig. 7. The exponential-like decay of the function denotes the strong influence of the two-coordinated C atoms on the transport.

**4. Conclusion**

We performed wave packet dynamical transport calculations for three models of graphene grain boundaries with increasing disorder: a periodic 5-7 grain boundary, a „serpentine" grain boundary, and a disordered one. The 5-7 boundary is composed of a periodic arrangement of 5-6-7 C rings, preserving the sp$^2$ lattice. The "serpentine" boundary lacks periodicity and is not rectilinear, still its transmission is similar to the 5-7 boundary. This is attributed to the very small number on broken $sp^2$ bonds and "extreme" (non 5-6-7 membered) polygons. The disordered grain boundary, however, contains numerous vacancies and "extreme" (4 or 8 membered) polygons. We identified these defects as the main scattering centres inside the disordered boundaries, which considerably diminish the electron transmission of the disordered grain boundaries. Hence, "defect engineering" may open a new way for development of high mobility graphene-based nanoelectronics.

**Acknowledgment**

The work in Hungary was supported in part by OTKA grant K101599 and an EU Marie Curie International Research Staff Exchange Scheme Fellowship within the 7th European Community Framework Programme (MC-IRSES proposal 318617 FAEMCAR project), the collaboration of Korean and Hungarian scientists was supported by the KRCF in the framework of the Korean-Hungarian Joint Laboratory for Nanosciences, the collaboration of Belgian and Hungarian scientists was supported by the bilateral agreement of the FNRS and HAS.

**Figures**

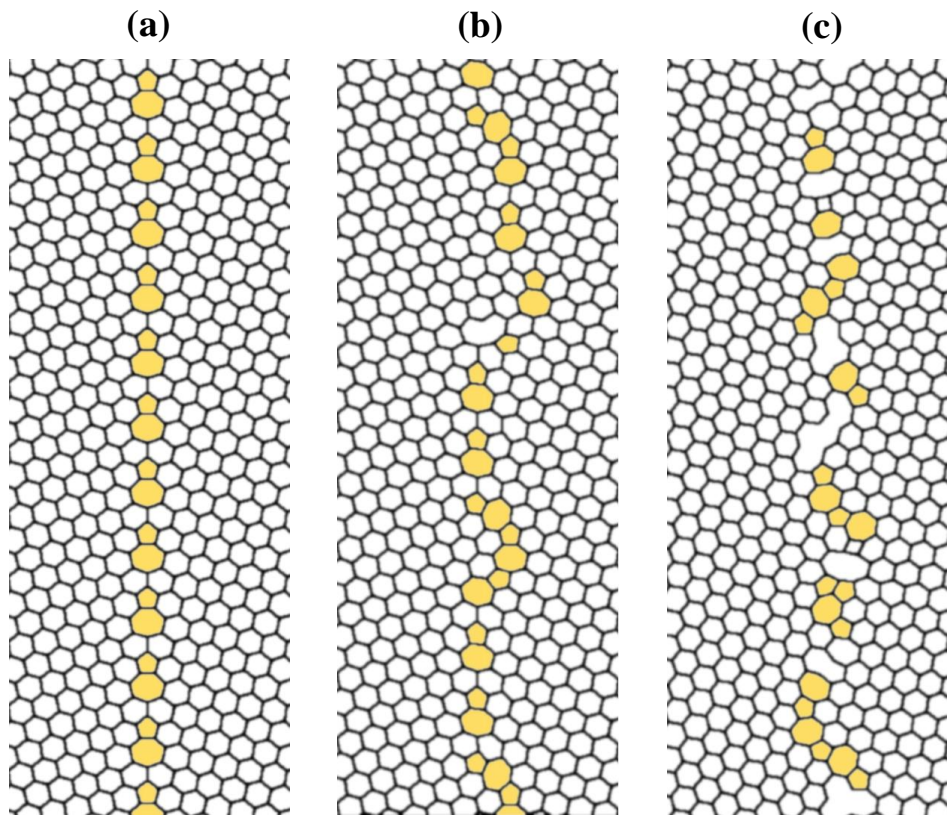

Fig. 1. Structures of the GBs, pentagons and heptagons are highlighted. (a) Periodical pentagon-heptagon GB (b) Serpentine GB without periodicity (c) Disordered GB containing several defects and non-hexagonal carbon rings. Structure (b) and (c) generated by a Monte-Carlo like procedure. See the text for details.

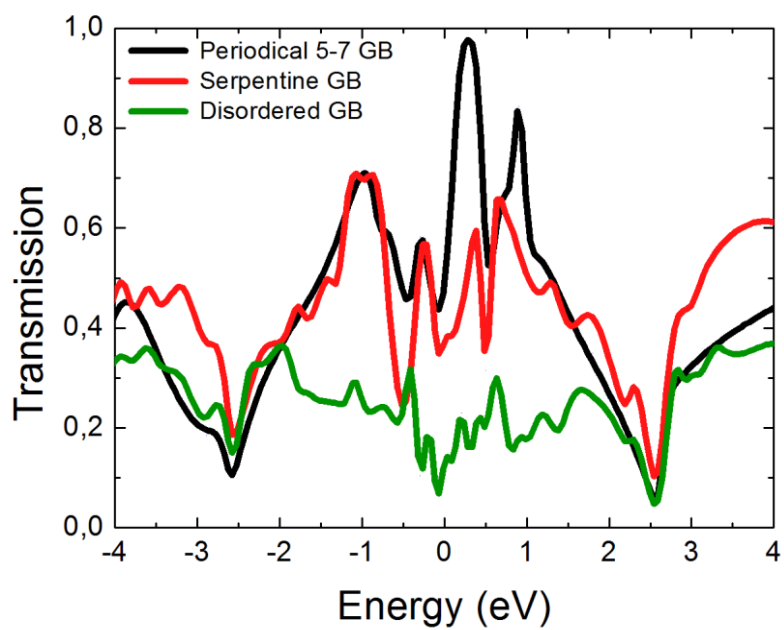

Fig. 2. Transmission functions of the GBs. The similar energy positions of the peaks in the case of the periodical 5-7 (black line) and serpentine GB (red line) originated from the pentagons and heptagons. Reduced transmission values appear for the low energy charge carriers in the case of the disordered GB (green line).

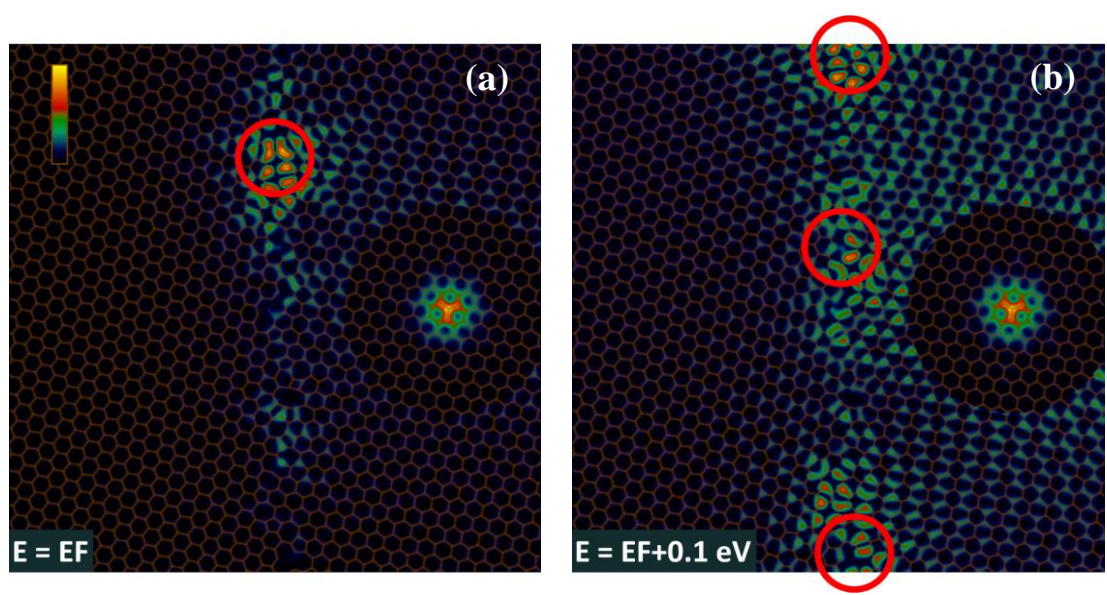

Fig. 3. $\rho(\vec{r}, E)$ probability density on the graphene sheet with the disordered GB around the Fermi energy as a color coded 2D (XY) section. We used a nonlinear color scale [see the scale bar in (a)] and different normalization below the STM tip inside and outside a circle with 1.5nm radius on the right-hand side grain. Note the strong localization around (a) the four membered carbon ring and (b) the vacancy type of defects marked by red circles. Size of the presentation window is 7.68 nm in both x and y directions.

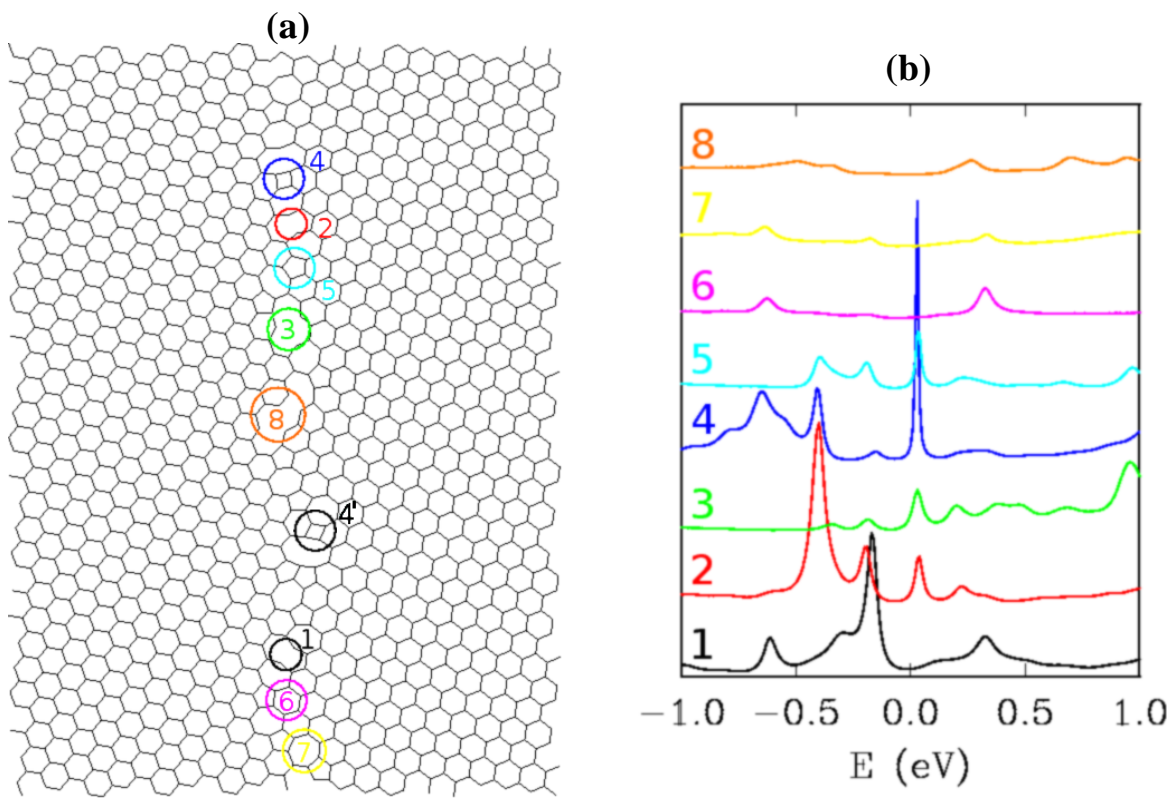

Fig. 4. Tight-binding electronic structure calculations for the disordered GB. (a) Different defects are numbered and marked by circles. (b) LDOS functions of the atoms within the circles.

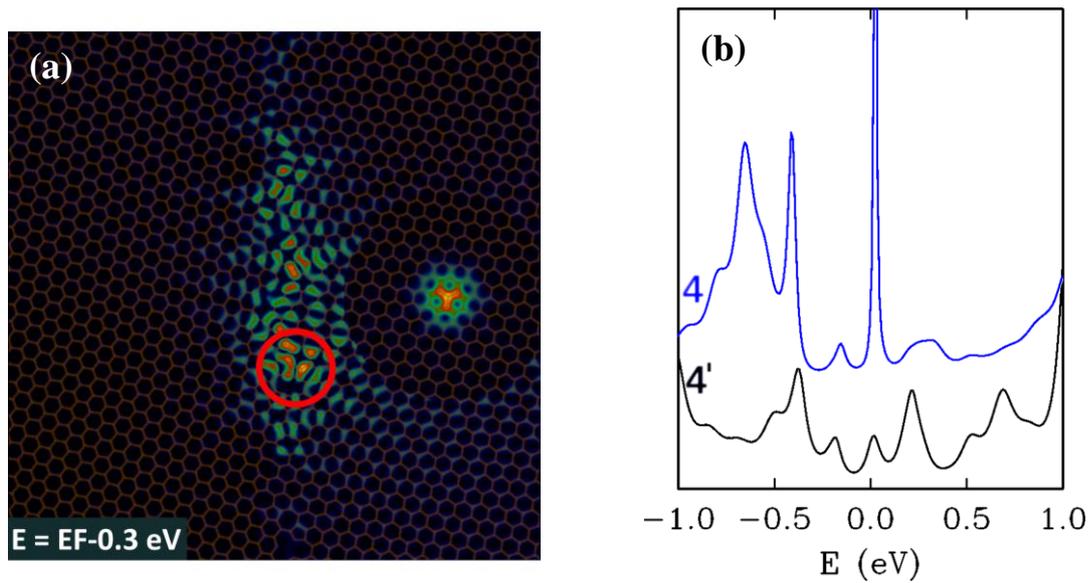

Fig. 5. (a) $\rho(\vec{r}, E)$ probability density on the graphene sheet at $E = EF - 0.3\,\text{eV}$. The four membered carbon ring nr 4'. (see Fig.4) marked by red circles. (b) LDOS functions on the two four membered carbon rings nr 4. and nr 4'.

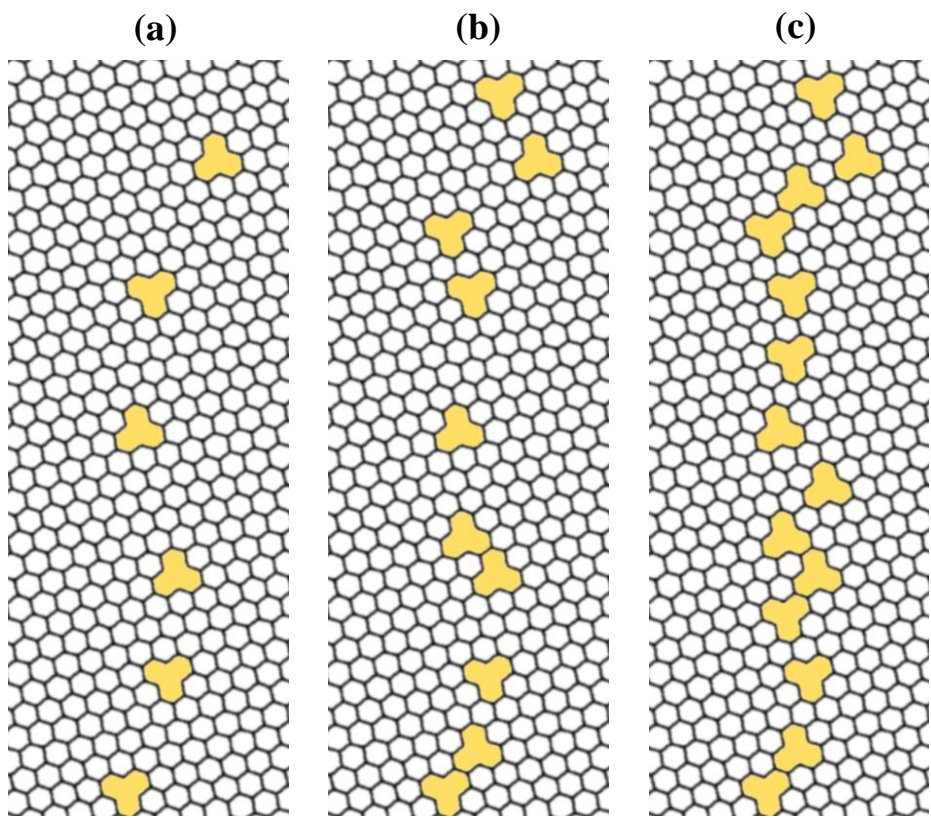

Fig. 6. Structures of graphene with increasing vacancy concentration along a curved line.

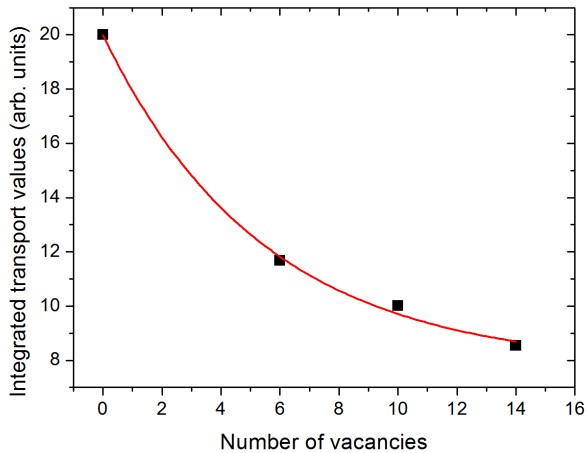

Fig. 7. Integrated transport values in the energy range $E = EF \pm 1\,\text{eV}$ for "vacancy-only" line defects as the function of the number of vacancies. Exponential-like decay of the transmission values highlights the effect of vacancies on the transport.

**References**

[1] N.M.R. Peres, Colloquium: The transport properties of graphene: An introduction, Reviews of Modern Physics 82 (2010) 2673-2700.

[2] A. Lherbier, S.M.-M. Dubois, X. Declerck, S. Roche, Y.-M. Niquet, J.-C. Charlier, Two-dimensional graphene with structural defects: elastic mean free path, minimum conductivity, and Anderson transition, Physical Review Letters 106 (2011) 046803.

[3] T.M. Radchenko, A.A. Shylau, I.V. Zozoulenko, Influence of correlated impurities on conductivity of graphene sheets: Time-dependent real-space Kubo approach, Physical Review B 86 (2012) 035418.

[4] M. Monteverde, et al., Transport and elastic scattering times as probes of the nature of impurity scattering in single-layer and bilayer graphene, Physical Review Letters 104 (2010) 126801.

[5] Z.H. Ni, et al., On resonant scatterers as a factor limiting carrier mobility in graphene, Nano Letters 10 (2010) 3868-3872.

[6] X. Li, et al., Large-area synthesis of high-quality and uniform graphene films on copper foils, Science 324 (2009) 1312–1314.

[7] C. Hwang, K. Yoo, S.J. Kim, E.K. Seo, H. Yu, L.P. Biró, Initial Stage of Graphene Growth on a Cu Substrate, The Journal of Physical Chemistry C 115 (2011) 22369–22374.

[8] K.S. Novoselov, V.I. Fal'ko, L. Colombo, P.R. Gellert, M.G. Schwab, K. Kim, A roadmap for graphene, Nature 490 (2012) 192–200.

[9] S. Bae, et al., Roll-to-roll production of 30-inch graphene films for transparent electrodes, Nature Nanotechnology 5 (2010) 574–578.

[10] P. Nemes-Incze, K.J. Yoo, L. Tapasztó, G. Dobrik, J. Lábár, Z.E. Horváth, C. Hwang, L. P. Biró, Revealing the grain structure of graphene grown by chemical vapor deposition, Applied Physics Letters 99 (2011) 023104.

[11] L. Tapasztó, P. Nemes-Incze, G. Dobrik, K.J. Yoo, C. Hwang, L.P. Biró, Mapping the electronic properties of individual graphene grain boundaries, Applied Physics Letters 100 (2012) 053114.

[12] L.P. Biró, Ph. Lambin, Grain boundaries in graphene grown by chemical vapor deposition, New Journal of Physics 15 (2013) 035024.

[13] O.V. Yazyev, S.G. Louie, Electronic transport in polycrystalline graphene, Nature Materials 9 (2010) 806–809.

[14] J. Kotakoski, J.C. Meyer, Mechanical properties of polycrystalline graphene based on a realistic atomistic model, Physical Review B 85 (2012) 195447.

[15] D.V. Tuan, J. Kotakoski, T. Louvet, F. Ortmann, J.C. Meyer, S. Roche, Scaling properties of charge transport in polycrystalline graphene, Nano Letters 13 (2013) 1730-1735.

[16] A.W.Tsen, et al., Tailoring electrical transport across grain boundaries in polycrystalline graphene, Science 336 (2012) 1143–1146.

[17] G.I. Márk, L.P. Biró, Ph. Lambin, Calculation of axial charge spreading in carbon nanotubes and nanotube Y junctions during STM measurement, Physical Review B 70 (2004) 115423.

[18] P. Vancsó, G.I. Márk, Ph. Lambin, C. Hwang, L.P. Biró, Time and energy dependent dynamics of the STM tip- graphene system, European Physical Journal B 85 (2012) 142.

[19] G.I. Márk, P. Vancsó, C. Hwang, Ph. Lambin, L.P. Biró, Anisotropic dynamics of charge carriers in graphene, Physical Review B 85 (2012) 125443.

[20] B.M. Garraway, K.A. Suominen, Wave-packet dynamics: new physics and chemistry in femto-time, Reports on Progress in Physics 58 (1995) 365–419.

[21] P. Vancsó, G.I. Márk, Ph. Lambin, A. Mayer, Y.-S Kim. C. Hwang, L.P. Biró, Electronic transport through ordered and disordered graphene grain boundaries, Carbon 64 (2013) 101–110.

[22] J.A. Fleck Jr., J.R. Morris, M.D. Feit, Time-dependent propagation of high energy laser beams through the atmosphere, Applied Physics 10 (1976) 129-160.

[23] A. Mayer, Band structure and transport properties of carbon nanotubes using a local pseudopotential and a transfer-matrix technique, Carbon 42 (2004) 2057–2066.

[24] M.D. Feit, J.A. Fleck, A. Steiger, Solution of the Schrödinger Equation by a Spectral Method, Journal of Computational Physics 47 (1982) 412-433.

[25] P. Simonis, C. Goffaux, P.A. Thiry, L.P. Biró, Ph. Lambin, V. Meunier, STM study of a grain boundary in graphite, Surface Science 511 (2002) 319–322.

[26] S. Kurasch, J. Kotakoski, O. Lehtinen, V. Skákalová, J. Smet, C.E. Krill, A.V. Krasheninnikov, U. Kaiser, Atom-by-Atom Observation of Grain Boundary Migration in Graphene, Nano Letters 12 (2012) 3168-3173.

[27] R. Haydock, V. Heine, M.J. Kelly, Electronic structure based on the local atomic environment for tight-binding bands. II, Journal of Physics C: Solid State 8 (1975) 2591–2605.

[28] H. Suzuura, T. Ando, Crossover from symplectic to orthogonal class in a two-dimensional honeycomb lattice, Physical Review Letters 89 (2002) 266603.

[29] S. Malola, H. Häkkinen, P. Koskinen, Structural, chemical, and dynamical trends in graphene grain boundaries, Physical Review B 81 (2010) 165447.